\documentclass[10pt, oneside, a4paper]{article}
\usepackage[titletoc]{appendix}
\usepackage{amsmath, amsthm, amssymb, amsfonts}
\usepackage{geometry}
\usepackage{graphicx}
\usepackage{natbib}
\usepackage{setspace}
\usepackage{longtable}
\usepackage{caption}
\usepackage{float}
\usepackage{enumerate}
\usepackage{url}

\theoremstyle{definition}

\newcommand{\renum}[1]{}
\input{tcilatex}

\begin{document}

\title{Comments on B. Hansen's Reply to "A Comment on: `A Modern
Gauss-Markov Theorem'", and Some Related Discussion\thanks{%
Address correspondence to Benedikt P\"{o}tscher, Department of Statistics,
University of Vienna, A-1090 Oskar-Morgenstern Platz 1. E-Mail:
benedikt.poetscher@univie.ac.at. }}
\author{Benedikt M. P\"{o}tscher \\
Department of Statistics, University of Vienna}
\date{June 2024\\
}
\maketitle

\begin{abstract}
In P\"{o}tscher and Preinerstorfer (2022) and in the abridged version P\"{o}%
tscher and Preinerstorfer (2024, published in Econometrica) we have tried to
clear up the confusion introduced in Hansen (2022a) and in the earlier
versions Hansen (2021a,b). Unfortunatelly, Hansen's (2024) reply to P\"{o}%
tscher and Preinerstorfer (2024) further adds to the confusion. While we are
already somewhat tired of the matter, for the sake of the econometrics
community we feel compelled to provide clarification. We also add a comment
on Portnoy (2023), a "correction" to Portnoy (2022), as well as on Lei and
Wooldridge (2022).
\end{abstract}

\section{Comments on Hansen (2024)}

We organize our comments by section in Hansen (2024). The more important
comments relate to Sections 2 and 3 in Hansen (2024).

\subsection{Comments on Section 1 in Hansen (2024)}

\begin{itemize}
\item Hansen wants to see Theorem 3.4 in P\"{o}tscher and Preinerstorfer
(2022, 2024) as providing an alternative proof of Theorem 4 in Hansen
(2022a), his `modern Aitken Theorem'. We rather see it as a result showing
that his `modern Aitken Theorem', i.e., Theorem 4 in Hansen (2022a), is
nothing else than the classical Aitken Theorem in disguise.

\item The reference to Portnoy (2022) is not entirely to the point: As
pointed out in Remark 3.6 in P\"{o}tscher and Preinerstorfer (2022) as well
as in Remark 3.6 in P\"{o}tscher and Preinerstorfer (2024), Portnoy (2022)
does \textbf{not} establish linearity of the estimators, but only Lebesgue
a.e. linearity. Why this difference matters is explained in the before
mentioned remarks. [This difference actually seems to have inspired Portnoy
to write the "correction" Portnoy (2023), see the discussion in Section 2
further below.]
\end{itemize}

\subsection{Comments on Section 2 in Hansen (2024)}

\begin{itemize}
\item The statement that P\"{o}tscher and Preinerstorfer (2024) does not
examine the case of independent sampling is misleading to say the least:
Hansen's (2022) results on independent sampling are discussed in the
abstract, in the introduction, and in the conclusion of P\"{o}tscher and
Preinerstorfer (2024), and the reader is referred to the more extensive
discussion in P\"{o}tscher and Preinerstorfer (2022), a paper Hansen is well
aware of. [This more extensive discussion was removed from P\"{o}tscher and
Preinerstorfer (2024) on the request of the editor of Econometrica.]

\item The discussion up to and including Theorem 5 (and the paragraph
following this theorem) is just a \textbf{repetition} of material in Hansen
(2022), and our criticism of this material given in Section 5 of P\"{o}%
tscher and Preinerstorfer (2022) still stands.

\item Theorem 11.1 is attributed to the book Hansen (2022b). The discussion
in this book claims that it is taken from Hansen (2022a). However, this
result is \textbf{nowhere} to be found in Hansen (2022a) (nor is it
contained in any of the earlier versions Hansen (2021a,b)): While Hansen
(2022a) indeed has a result for the location case (Theorem 7), this is a 
\textbf{different} result. [Hansen (2021a,b) also has results for the
location case (Theorem 6 in both versions), but again these are results 
\textbf{different} from Theorem 11.1 in Hansen (2022b). As discussed in P%
\"{o}tscher and Preinerstorfer (2022), one of these two results is correct,
the other one is incorrect.]

\item A sketch of a proof of Theorem 11.1 is given in the book Hansen
(2022b). This sketch is modelled after proofs in Hansen (2022a).
Unfortunately, this sketch of a proof rests on Theorem 10.6 in Hansen
(2022b), his rendition of the Cram\'{e}r-Rao lower bound, a rendition that
is lacking mathematical rigor and likely is not correct as given, because
well-known regularity conditions (e.g., conditions for interchange of
integration and differentiation) are omitted from Theorem 10.6 in Hansen
(2022b). As a consequence of using Theorem 10.6 in Hansen (2022b), such
regularity conditions are then \textbf{not} checked in the proof of Theorem
11.1. While the proof of Theorem 11.1 can perhaps be repaired, what is 
\textbf{given} in Hansen (2022b) certainly does \textbf{not} constitute a
proof of Theorem 11.1.\footnote{%
There is a further problem here: From the discussion surrounding Theorem
10.6 in Hansen (2022b) it appears that this theorem is given for parametric
models defined through densities w.r.t. Lebesgue measure. However, in a
rigorous proof of Theorem 11.1 one needs to be able to apply the Cram\'{e}%
r-Rao bound also in models that are not necessarily described by Lebesgue
densities, but by densities w.r.t. another base measure. This problem can
certainly be fixed, but Theorem 10.6 \textbf{as given} is not applicable.}

\item Fortunately, there is actually \textbf{no need} to come up with a
proof of Theorem 11.1 as it is an old result already proved in Halmos
(1946). This is well-known, and is also discussed in Section 6 of P\"{o}%
tscher and Preinerstorfer (2022), a discussion that can hardly have escaped
Hansen's attention.\footnote{%
The reference to Halmos (1946) is also mentioned in Appendix A of P\"{o}%
tscher and Preinerstorfer (2024).} Nevertheless the book Hansen (2022b) is
silent on this issue and does not make any mention of Halmos (1946).

\item The claim that Theorem 11.1 would be a strict improvement over the Cram%
\'{e}r-Rao Theorem is obviously nonsense as the Cram\'{e}r-Rao Theorem
applies to unbiased estimators in general parametric models (the estimators
and the model satisfying certain regularity conditions). Maybe Hansen wanted
to say that Theorem 11.1 is a strict improvement over the result that in the
context of i.i.d. normal data the mean is best unbiased (which can be
obtained from the Cram\'{e}r-Rao Theorem). But this again is nonsense, as
the classes of unbiased estimators in these two statements are not
guaranteed to be the same.

\item It is interesting to note that the same problems that plague the proof
of Theorem 11.1 in Hansen (2022b) actually also plague the proofs of
Theorems 4, 5, 6, and 7 in Hansen (2022a) as they are again based on the
non-rigorous rendition of the Cram\'{e}r-Rao lower bound given in Theorem
10.6 in Hansen (2022b). While these proofs can possibly be repaired, we have
not bothered to check this in any detail.

\item We note that the set $\boldsymbol{F}_{2}^{0}$ defined in Hansen (2024)
is \textbf{different} from the set $\boldsymbol{F}_{2}^{0}$ defined in
Hansen (2022a), and both are different from the set $\boldsymbol{F}_{2}^{0}$
defined in Hansen (2021a,b). This is unfortunate as it creates unnecessary
confusion.

\item An earlier version (November 2023) of Hansen (2024) claimed that our
counterexample given in Appendix A of P\"{o}tscher and Preinerstorfer (2024)
(also given in Appendix A of P\"{o}tscher and Preinerstorfer (2022)) is
mislabeled as a "counterexample". This is nonsense, and fortunately is not
repeated in Hansen (2024).
\end{itemize}

\subsection{Comments on Section 3 in Hansen (2024)}

\begin{itemize}
\item Here Theorems 5 and 11.1 are repeated as Theorems 5' and 11.1' in a
less formal notation. This lends itself to an increased chance of
misunderstanding. For example, Theorem 5' in this section is not a precise
mathematical statement; it can easily be understood as a statement where the 
$\sigma _{i}^{2}$ are arbitrary but fixed, which then gives an incorrect
statement. We doubt that the two sentences following that theorem will help
much to avoid the possibility of misunderstanding. Especially so, when this
statement is specialized to the case of OLS. Here the typical reader will
believe that one can set $\sigma _{i}^{2}=\sigma ^{2}$ in that theorem. We
thus warn against using the imprecise formulation of the results given in
that section. That section should best be ignored.
\end{itemize}

\section{Comments on Portnoy (2023)}

In the "correction" Portnoy (2023) and the accompanying supplementary
material, Portnoy claims that the theorem in Portnoy (2022) would state
linearity of the estimators and that the proof would be in error. This is
not correct as claimed, since the theorem in Portnoy (2022) \textbf{only}
states Lebesgue a.e. linearity of the estimators considered, and \textbf{not}
linearity. [The proof in Portnoy (2022) rests on Lusin's Theorem and
inspection shows that it only delivers Lebesgue a.e. linearity. Whether this
proof is actually correct, we have not checked.] It seems that in the
"correction" Portnoy (2023) now wants to give a proof of a full linearity
result in hindsight after such a result has appeared in Theorem 3.4 in P\"{o}%
tscher and Preinerstorfer (2022) (which does establish linearity by
exploiting unbiasedness under certain discrete distributions). It is
welcomed that Portnoy now provides an alternative proof of the full
linearity result, but this is an \textbf{extension} of Portnoy (2022) rather
than a \textbf{correction}.\footnote{%
Again, we have not checked the details of the proof given in Portnoy (2023)
(and the supplementary material).} Why this is advertised as a "correction"
to Portnoy (2022) puzzles us. Also note that, for some reason, Portnoy
(2023), and the supplementary material, do \textbf{not} acknowledge the full
linearity result (Theorem 3.4) in P\"{o}tscher and Preinerstorfer (2022) at
all.\footnote{%
This full linearity result is also given in the abridged version P\"{o}%
tscher and Preinerstorfer (2024).}

\section{Comments on Lei and Wooldridge (2022)}

Lei and Wooldridge (2022) incorrectly claim to have given a proof of the
linearity result on Twitter, and they provide Twitter links in their paper.
Inspection of the links shows that what is discussed there is a conjecture
that the estimators in question must be linear, but there is no proof given
there at all for at least two reasons: (i) The purported proof claims that
the set of $n\times n$ symmetric matrices spans $\mathbb{R}^{n\times n}$, a
blatantly false claim. (ii) It also rests on Theorem 4.3 in Koopmann (1982),
for which no complete proof is given in that reference, see the discussion
on p.9 of P\"{o}tscher and Preinerstorfer (2022) (see also Remark 3.7 and
Appendix B in P\"{o}tscher and Preinerstorfer (2024)). Hence, while Lei and
Wooldridge correctly guessed linearity (like probably many others), they did 
\textbf{not} provide a proof in their Twitter discussion. [Lei and
Wooldridge (2022), released almost a year after their Twitter discussion,
now claims to have finally managed to come up with a proof of Theorem 4.3 of
Koopmann (1982), which -- if correct -- would void objection (ii), but not
(i).]

P\"{o}tscher and Preinerstorfer (2022) provided a proof of the linearity
result (Theorem 3.4) that does \textbf{not} rest on Theorem 4.3 in Koopmann
(1982). Additionally, P\"{o}tscher and Preinerstorfer (2022) also provided
(another) proof \textbf{conditional} on Theorem 4.3 in Koopmann (1982),
explicitly stressing the conditionality. Lei and Wooldridge (2022) further
claim that this conditional proof in P\"{o}tscher and Preinerstorfer (2022)
would be identical to their own purported proof on Twitter. This is \textbf{%
incorrect}, as the conditional proof in P\"{o}tscher and Preinerstorfer
(2022) does \textbf{not} rely on the incorrect claim (i) appearing in Lei's
and Wooldridge's Twitter discussion; furthermore, -- contrary to the
purported Twitter-proof -- the conditional proof in P\"{o}tscher and
Preinerstorfer (2022) is correct.

We abstain from providing a list of all the errors and inconsistencies that
can be found in Lei and Wooldridge (2022), and that make it difficult to
gauge the actual contribution of that paper.

\section{References}

\ \ \ \ Hansen, Bruce E. (2021a): "A Modern Gauss-Markov Theorem." September
2021. Version accepted for publication in Econometrica.

Hansen, Bruce E. (2021b): "A Modern Gauss-Markov Theorem." December 2021.
Update of September 2021 version accepted for publication in Econometrica.

Hansen, Bruce E. (2022a): "A Modern Gauss-Markov Theorem." \emph{Econometrica%
} 90, 1283-1294.

Hansen, Bruce E. (2022b): \emph{Probability and Statistics for Economists},
Princeton University Press.

Hansen, Bruce E. (2024): "Reply to: "A Comment on: "A Modern Gauss-Markov
Theorem""" \emph{Econometrica} 92, 925-928.

Lei, L. and J. Wooldridge (2022): "What Estimators are Unbiased for Linear
Models?", arXiv:2212.14185.

Koopmann, R. (1982): \emph{Parametersch\"{a}tzung bei a priori Information}.
Vandenhoeck\&Ruprecht, G\"{o}ttingen.

Portnoy, S. (2022): "Linearity of Unbiased Linear Models Estimators." \emph{%
The American Statistician} 76, 372-375.

Portnoy, S. (2023): \ "Correction to 'Linearity of Unbiased Linear Models
Estimators.'" \emph{The American Statistician} 77, 237.

P\"{o}tscher, B.M. \& D. Preinerstorfer (2022): "A Modern Gauss-Markov
Theorem? Really?" arXiv:2203.01425v5.

P\"{o}tscher, B.M. \& D. Preinerstorfer (2024): "A Comment on: "A Modern
Gauss-Markov Theorem"" \emph{Econometrica}, 92, 913-924.

\end{document}